\documentclass[aps,prl,twocolumn,showpacs,superscriptaddress]{revtex4}
\usepackage[latin1]{inputenc}
\usepackage[english]{babel}
\usepackage{bm,graphicx,graphics,amsmath,amssymb,epsfig,color}

\usepackage[colorlinks,citecolor=blue,linkcolor=cyan]{hyperref}
\usepackage[usenames,dvipsnames,svgnames,table]{xcolor}

\usepackage{mathptmx}

\renewcommand{\section}[1]{{\par\it #1.---}\ignorespaces}

\def \be {\begin{equation}}
\def \ee {\end{equation}}
\def \beA {\begin{eqnarray}}
\def \eeA {\end{eqnarray}}

\def \average#1{\left\langle #1 \right\rangle}

\begin{document}
\title{Coherent energy transport in classical nonlinear oscillators: an analogy with the Josephson effect}

\author{Simone Borlenghi} 
\affiliation{Department of Physics and Astronomy, Uppsala University, Box 516, SE-75120 Uppsala, Sweden.}
\affiliation{Department of Materials and Nanophysics,  School of Information and Communication Technology, \\Electrum 229, Royal Institute of Technology, SE-16440 Kista, Sweden.}
\author{Stefano Iubini}
\affiliation{Centre de Biophysique Moléculaire (CBM), CNRS-UPR 4301 Rue Charles Sadron, F-45071 Orléans, France}
\author{Stefano Lepri}
\affiliation{Istituto dei Sistemi Complessi, Consiglio Nazionale delle Ricerche, Unit\`{a} Operativa di Firenze, Via Madonna del Piano 10 I-50019 Sesto Fiorentino, Italy.} 
\author{Lars Bergqvist} 
\affiliation{Department of Materials and Nanophysics,  School of Information and Communication Technology, \\Electrum 229, Royal Institute of Technology, SE-16440 Kista, Sweden.}
\author{Anna Delin}
\affiliation{Department of Physics and Astronomy, Uppsala University, Box 516, SE-75120 Uppsala, Sweden.}
\affiliation{Department of Materials and Nanophysics,  School of Information and Communication Technology, \\Electrum 229, Royal Institute of Technology, SE-16440 Kista, Sweden.}
\author{Jonas Fransson}
\affiliation{Department of Physics and Astronomy, Uppsala University, Box 516, SE-75120 Uppsala, Sweden.}

\begin{abstract}
By means of a simple theoretical model and numerical simulations, we demonstrate the presence of persistent energy currents in a lattice of classical nonlinear oscillators
with uniform temperature and chemical potential. In analogy with the well known Josephson effect, the currents are proportional to the sine of the phase differences between the oscillators. 
Our results elucidate a general aspects of non-equilibrium thermodynamics and point towards novel way to practically control transport phenomena in a large class of systems.
We apply the model to describe the phase-controlled spin-wave current in a bilayer nano-pillar. 
\end{abstract}

\maketitle

Energy transport is common to all physical systems out of thermal equilibrium. The standard approach to non-equilibrium thermodynamics describes transport in terms of generalised forces and coupled currents
\cite{onsager31a,onsager31b,kubo57a,kubo57b}, a typical example being the Fourier law, that relates temperature gradient to the heat flux \cite{lepri03,dhar08}. 

In this Letter, we demonstrate the presence of persistent energy currents in a network of classical nonlinear oscillators with uniform temperature and chemical potential.
In strong analogy with the well known Josephson effect \cite{josephson62}, the currents are generated only by the phase differences between the oscillators. 
In fact, the phases play the role of additional thermodynamical forces, that drive the system out of equilibrium in a way similar to a non-uniform temperature distribution.

In the quantum regime, phase-controlled heat transfer and rectification has been already observed in superconducting Josephson junctions \cite{maki65,giazotto12,perez13,perez14}. So far, however, the Josephson effect has been considered a purely quantum phenomenon that belongs to the realm of superconductivity and Bose-Einsten condensates (BEC). By contrast, in this Letter we show that the main aspects of the Josephson  physics are common to a large class of \emph{classical} systems out of thermal equilibrium, as a direct consequence of the fluctuation-dissipation theorem. 

We start by a general description of the model and numerical simulations of energy transport between coupled oscillators. 
Then, we will apply the model to the micromagnetic study of the phase-controlled spin-transport in a Permalloy (Py) bilayer nano-pillar.

Let us consider the following model 
\begin{align}
i\dot{\psi}_m = & \omega_m(p_m)\psi_m-i\Gamma_m(p_m)\psi_m+i\mu_m\psi_m
\nonumber\\&
	+\sum_{n} A_{mn}\psi_n+ i\xi_m(t),
\label{eq:dnls}
\end{align}
which generalises of the well known discrete nonlinear Schr\"odinger equation (DNLS) to a network of coupled nonlinear oscillators with arbitrary geometry and local 
complex amplitudes $\psi_m=\sqrt{p_m(t)}{\rm{e}}^{i\phi_m(t)}$. 

 Eq.(\ref{eq:dnls}), finds application in many branches of physics, such as Bose-Einstein condensates, photonics waveguides, lasers and mechanical oscillators \cite{kevrekidis01,eilbeck03,rumpf03}. 
In fact, irrespective of the underlying physical process, the small amplitude dynamics of a large class of nonlinear oscillators can be written in the DNLS form \cite{johansson04,johansson06}, 
which is usually referred to as the universal nonlinear oscillator model \cite{slavin09}.
In particular, the DNLS is used to model the dynamics of several types of classical spin systems \cite{borlenghi14a,borlenghi14b,slavin09},

\begin{figure}
\begin{center}
\includegraphics[width=0.99\columnwidth]{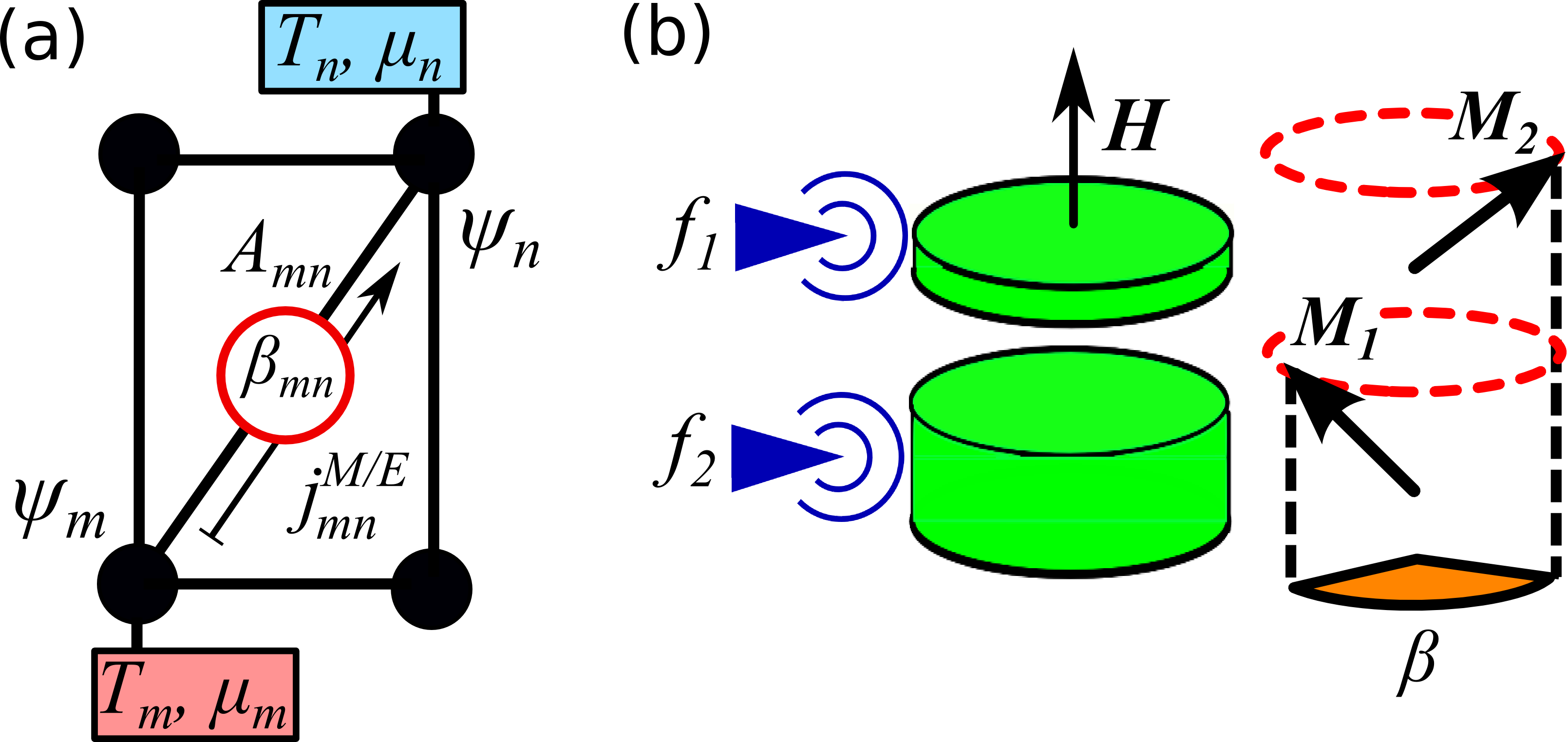}
\end{center}
\caption{(Color online) (a) Oscillator network, where the oscillators $(\psi_m,\psi_n)$ (represented by the dots) are coupled through the complex matrix $A_{mn}=C_{mn}\rm{e}^{i\beta_{mn}}$ (represented by the links). 
Thermal baths at different temperatures and chemical potentials $(T_m,\mu_m)$ together with the phases $\beta_{mn}$ can be used to drive the system out of equilibrium 
and control the flows $j_{mn}^{M/E}$ between the nodes $(m,n)$. (b) Double disk nano-pillar studied in the micromagnetic simulations. The system behaves as two coupled macrospins $\bm{M}_{1,2}$, precessing around the effective field $\bm{H}$.
The dynamics is driven by rf fields $f_{1,2}$, which control the phase mismatch $\beta$.} 
\label{fig:figure1}
\end{figure}

The first term on the right hand side of Eq.(\ref{eq:dnls}) is the nonlinear frequency, 
The second term $\Gamma_m(p_m)=\alpha_m\omega_m(p_m)$ is the nonlinear damping, proportional to the parameter $\alpha_m$. 
In spin systems those terms are proportional to the local applied field, and the nonlinearity is modeled as an expansion in powers of $p_n$ \cite{slavin09,borlenghi14b}.

Next comes the chemical potential $\mu_m$, which acts as a torque that controls the relaxation time towards the reservoirs by compensating the damping \cite{slavin09,iubini13,borlenghi14b}. 
In magnonics and spintronics devices, this can be controlled through spin transfer torque \cite{slavin09,borlenghi14b}.

The matrix $A_{mn}=C_{mn}{\rm{e}}^{i\beta_{mn}}$, which accounts for the coupling and specifies the geometry of the system, may depend on different mechanisms and includes both long and short range (nearest neighbour) interactions. 
The phase $\beta_{mn}$ also depends on the geometry and on the coupling mechanism. Note that $A_{mn}$ is hermitian in the absence of dissipation, while with dissipation one normally has $\beta_{mn}=\beta_{nm}$, as it will be clarified below. 
We will consider for simplicity a linear coupling, but our analysis can be extended without difficulty to more general cases. 

The last term of Eq.(\ref{eq:dnls}) describes thermal fluctuations in terms of a complex Gaussian random variable with the usual statistical properties $\average{\xi_m(t)}=0$ and $\average{\xi_m^*(t)\xi_n(t^\prime)}=D_m(p_m)T_m\delta_{mn}\delta(t-t^\prime)$ \cite{slavin09,iubini13,borlenghi14b}. The nonlinear diffusion constant $D_m(p_m)=\lambda\Gamma_m(p_m)/\omega_m(p_m)$ specifies the coupling strength with the underlying phonon bath, as prescribed by the fluctuation-dissipation
theorem \cite{iubini13,slavin09}. $T_m$ is the bath temperature, while $\lambda$ is a parameter that depends on the geometry of the oscillators \cite{slavin09}.

Eq.(\ref{eq:dnls}) is conveniently written in the phase-amplitude representation as \cite{slavin09,iubini13,borlenghi14a,borlenghi14b}. 
\begin{align}
\dot{p}_m = & 
	2[\mu_m-\Gamma_m(p_m)]p_m+2\sqrt{p_m}{\rm{Re}}[\tilde{\xi}_m(t)]
\nonumber\\&
        +2D_m(p_m)T_m+\sum_n j^M_{mn}
\label{eq:amplitude}
\\
\dot{\phi}_m = &
	-\omega_m(p_m)+\frac{1}{\sqrt{p_m}}{\rm{Im}}[\tilde{\xi}_m(t)]
\nonumber\\&
       - \sum_{n}K_{mn}\cos(\phi_m-\phi_n+\beta_{mn}).
\label{eq:phase}
\end{align}
Eq.(\ref{eq:amplitude}) is the continuity equation which relates the time evolution of the oscillator power $p_m=|\psi_m|^2$ to the currents 
$j_{mn}^M=2{\rm{Im}}[A_{mn}\psi_m^*\psi_n]$ between oscillators $(m,n)$. In spin systems, the oscillator power is related to the precession amplitude of the local magnetic moment
and $j_{mn}^M$ describes the transport of the component of the magnetisation along the precession axis \cite{slavin09,borlenghi14a,borlenghi14b}, as it will described below.
For convenience, we will call it "magnetisation current", even if it can describe other types of transport processes, such as the particle flow in BEC \cite{trombettoni01}.

The local chemical potentials $\mu_m$ and dampings $\Gamma_m(p_m)$ act respectively as sources and drains for $p_m$.
The constant term $2D_m(p_m)T_m$ comes from the coupling with the baths and ensures that the average values of $p_m$ is never zero at finite temperature \cite{borlenghi14b}. The two coupled currents $j_m^{M/E}$ can be controlled by a non-uniform distribution of spin temperature and chemical potential $(T_m,\mu_m)$, which represent the associated thermodynamical forces \cite{iubini12,iubini13}. The stochastic variable $\tilde{\xi}_m(t)=\xi_m(t){\rm{e}}^{-i\phi_m(t)}$ models a thermal bath with the same properties as $\xi_m(t)$ \cite{slavin09}. 

Eq.(\ref{eq:phase}) describes the dynamics of the local phase $\phi_m$ \cite{stratonovich67,pikovsky01}, which depends on the coupling
$K_{mn}=\sqrt{p_n/p_m}C_{mn}$ between oscillators $(m,n)$. When all the oscillators have the same coupling, Eq.(\ref{eq:phase}) reduces to the well known Kuramoto model \cite{kuramoto75}.
  
The conservative part of Eq.(\ref{eq:dnls}) is given by the functional derivative  $\dot{\psi}_m=-i{\delta\mathcal{H}}/{\delta\psi_m^*}$ of the total Hamiltonian  $\mathcal{H} = {\sum_{m,n}}[\omega_m(p_m)p_m+A_{mn}\psi_m^*\psi_n +c.c.]$ \cite{slavin09,iubini13}. Computing the time evolution of $\mathcal{H}$ gives the energy current $j_{mn}^E=2{\rm{Re}}[A_{mn}\dot{\psi}_m^*\psi_n]$ \cite{lepri03,iubini12,iubini13,borlenghi14a,borlenghi14b}. 

Using the phase-amplitude representation, the two currents are conveniently written as 
\begin{align}
j_{mn}^M =&
	2C_{mn}\sqrt{p_mp_n}\sin{(\phi_m-\phi_n+\beta_{mn})}
\label{eq:jm}
\\
j_{mn}^E =&
	-C_{mn}\dot{p}_m\sqrt{\frac{p_n}{p_m}}\cos(\phi_m-\phi_n+\beta_{mn})
\nonumber\\&
        +2C_{mn}\sqrt{p_mp_n}\omega_m(p_m)\sin(\phi_m-\phi_n+\beta_{mn}).
\label{eq:je}
\end{align}
In many situations (at zero temperature) the system reaches a steady state in which the condition $\dot{p}_m=0$ is satisfied \cite{slavin09,borlenghi14b}, such that only the second 
term of $j_{mn}^E$ survives. In this case both currents have a similar profile (up to a rescaling factor), as it will be shown below. 
The time and ensemble averaged currents have the natural interpretation of correlations between the oscillators $(m,n)$.
From Eqs.(\ref{eq:jm},\ref{eq:je}) one can see that, if the oscillators ($m,n$) are not phase locked $\phi_m(t)\neq\phi_n(t)$, so that the currents oscillate in time around zero and
they vanish in average. In this free running condition, there is \emph{no net transport} of energy and magnetisation through the system \cite{borlenghi14a,borlenghi14b}. 
On the contrary, when the oscillators are phase locked, $\phi_m(t)-\phi_n(t)\approx0$, the net currents are proportional to $\sin\beta_{mn}$. 

This result suggests that energy and magnetisation can flow between \emph{two sources with the same temperatures and chemical potentials} whenever the phase $\beta_{mn}$ is finite.
This situation is similar to the DC Josephson effect \cite{josephson62}, where the phase difference between the superconducting order parameters allows a persistent electrical current 
to propagate through a superconducting tunnel junctions, without any external bias.

The crucial observation in our case is that the system reaches thermal equilibrium (relaxing to the usual Gibbs distribution) only if the coupling has the form $A_{mn}=\mathcal{A}_{mn}[1-i\Gamma_m(p_m)]$, where $\mathcal{A}_{mn}$ is a real matrix \cite{iubini13}. 
This condition of dissipative coupling, prescribed by the fluctuation-dissipation theorem, amounts to fixing the imaginary part of $A_{mn}$ and consequently $\beta_{mn}$. 
Modifying this phase drives the system out of equilibrium, in a way similar to a non-uniform temperature or chemical potential. 

This fact is most intriguing since the Josephson effect has been considered so far as a purely quantum phenomenon.  
Normally quantum effects are lost in the classical limit, and a decisive role in this transition is played by the coupling with the environment, which brings decoherence through thermal fluctuations and irreversibility (dissipation) \cite{zurek03}. 

However, in the present case, fluctuations and dissipation are the key ingredients to observe this effect. In fact, the thermal baths ensure non-vanishing values of $p_m$ \cite{borlenghi14b},
while the phase $\beta_{mn}$ makes the phase-locking more robust. This can be see from Eq.(\ref{eq:phase}), where $\beta_{mn}$ 
controls the phase mismatch $\phi_m(t)-\phi_n(t)$ between oscillators $(m,n)$.

Fluctuations typically break the exact synchronisation and lead to phase diffusion \cite{borlenghi14b,stratonovich67}, so that this Josephson-like effect holds at low temperature.

The generality of the discussed model strongly suggests that this Josephson-like effect should be a general feature common to many physical systems, and not a peculiarity of superconductivity and quantum mechanics.
The main difference between the quantum and classical case consists in the nature of fluctuations and in the different distribution (respectively Bose and Gibbs) attained at thermal equilibrium \cite{biehs13}.
 

To substantiate the above arguments, we turn now to numerical simulations.
We consider a Schr\"odinger dimer, the simplest possible realisation of Eq.(\ref{eq:dnls}), consisting of only two coupled nonlinear oscillators. 
The two nonlinear frequencies and damping rates read respectively $\omega_n(p_n)=\omega^0_n\times(1+2p_n)$ and $\Gamma_n(p_n)=\alpha\omega_n(p_n)$, $n=1,2$ (in units where $k_B=1$).
The oscillators are connected to two Langevin baths with temperatures $T_n$ and coupling strength $D_1=D_2=2\alpha$. In all the simulations, the chemical potentials $\mu_n$ are set to zero, 
and we consider a symmetric coupling $A_{12}=A_{21}=C(1-i\alpha){\rm{e}}^{i\beta}$. The equations of motion for the dimer were solved numerically using a fourth order Runge-Kutta algorithm with a time step ${\rm{d}}t=10^{-3}$ model units. The other parameters, used in all our simulations are $C=0.1$ and $\alpha=0.02$. 
The relevant observables (local powers and currents) were time averaged in the stationary state over an interval of $2\times 10^6$ time steps and then ensemble-averaged over 100 samples with different realisations of the thermal field.

\begin{figure}
\begin{center}
\includegraphics[width=0.99\columnwidth]{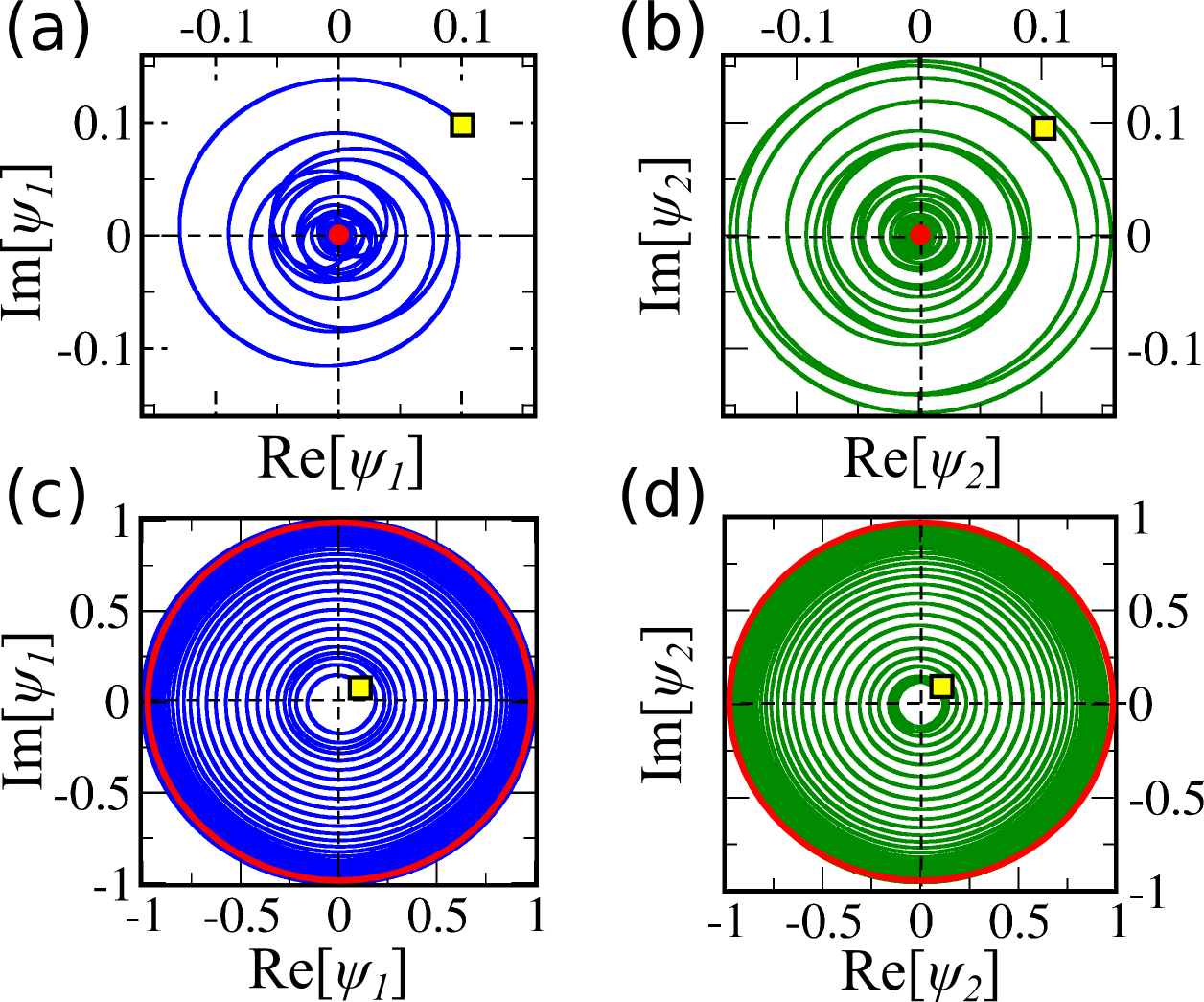}
\end{center}
\caption{(Color online) Phase portraits. Panels (a) and (b) show respectively the dynamics of oscillators 1 and 2, for $\beta=0$. In this case the motion is damped and phase space reduces to a single point (denoted by the red dots). 
Panels (c) and (d) show the case where $\beta=1.2\pi$, and the two oscillators reach a limit cycle at $|\psi_n|^2\approx 1$ (denoted by the red circles). The yellow squares indicate the initial condition.}
\label{fig:figure2}
\end{figure}

Fig.\ref{fig:figure2} shows the phase portrait of the two oscillators at zero temperature and frequencies $(\omega_1^0=1,\omega_2^0=1.2)$ , starting from the initial condition $\psi_1(0)=\psi_2(0)=0.1(1+i)$. When $\beta=0$ [panels (a) and (b)], the system is dissipative and the phase space collapses to a point.
However, when $\beta=1.2\pi$ [panels (c) and (d)], the imaginary part of the coupling changes sign and acts as a gain. In this condition, the two oscillators are self-sustained and reach a limit cycle close to $|\psi_n|^2\approx1$. 
This corroborates our model, showing that $\beta$ acts as a real torque that forces the dynamics.

\begin{figure}
\begin{center}
\includegraphics[width=0.99\columnwidth]{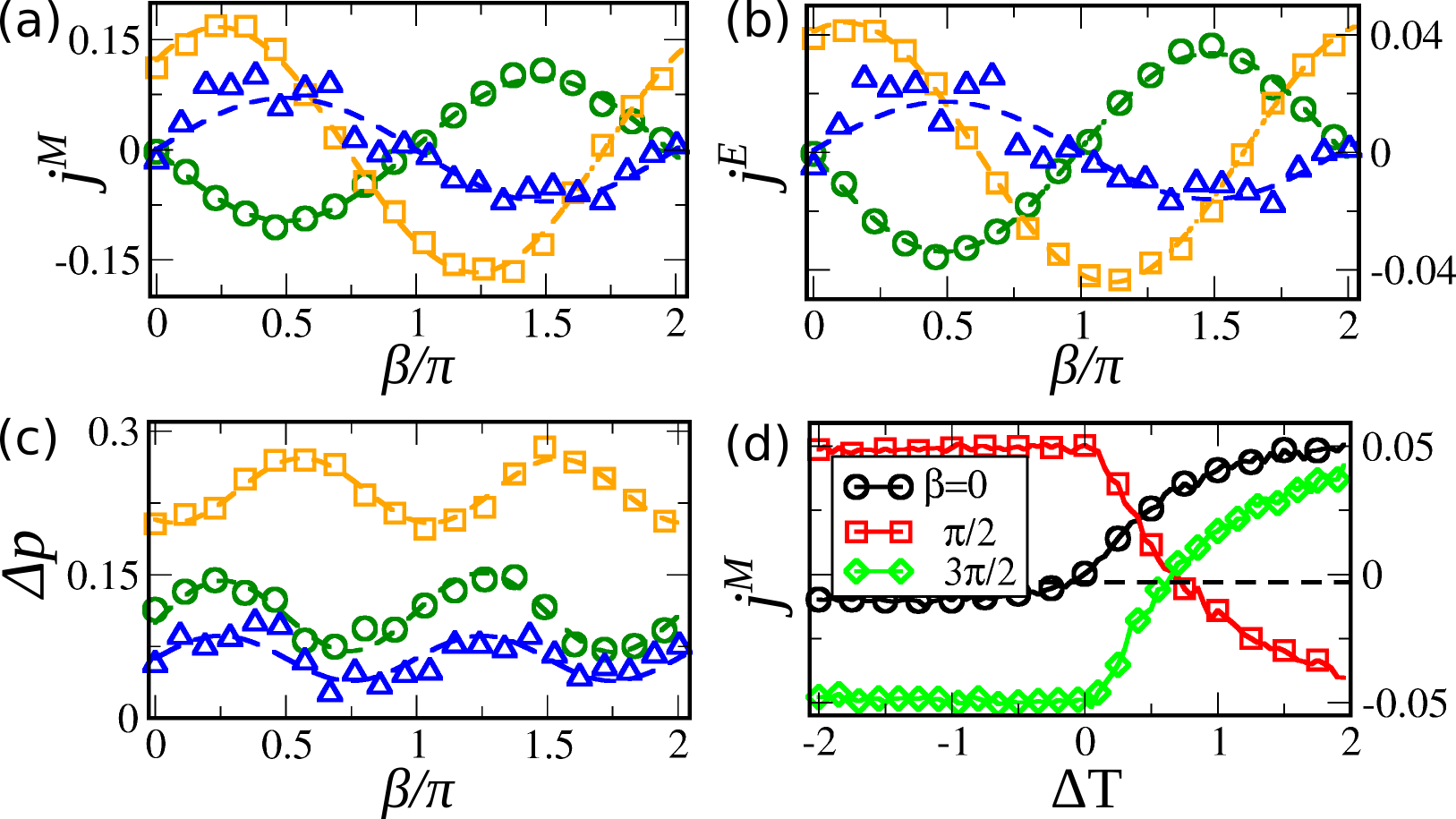}
\end{center}
\caption{(Color online) Josephson effect. Panels (a) and (b) display respectively energy and magnetisation currents, while (c) shows the power difference $\Delta p$ vs the coupling phase $\beta$. 
Different symbols correspond to different system parameters, see text. For better visibility, the figures in blue triangles are magnified by a factor 5 [panels (a) and (b)] and 2.5 (c), while
the curve in orange squares in panel (c) is shifted downwards of 0.25. The dashed lines are fit with sine functions. (d) Magnetization current $j^{M}$ vs temperature difference $\Delta T$, computed for different values of $\beta$. The lines are guide to the eye}
\label{fig:figure3}
\end{figure}

The transport properties at finite temperature are displayed in Fig.\ref{fig:figure3} (a) and (b), which show respectively energy and magnetisation currents as a function of $\beta$, for different values of the bath temperatures. Green circles 
and blue triangles show the case where the two baths have the same temperatures $T_1=T_2=0.8$, and frequencies are respectively $(\omega^0_1=1,\omega^0_2=2)$ and $(\omega^0_1=1,\omega^0_2=1.2)$. 
As discussed in the model, both currents are proportional to $\sin\beta$ and have similar profiles. Note in particular that the currents vanish when $\beta=0$. In this case the coupling has the correct form that respects the fluctuation-dissipation theorem and allows the system to reach thermal equilibrium.

The orange squares correspond the case where $T_1=1.2$ and $T_2=0.2$ and $(\omega^0_1=1,\omega^0_2=2)$. In this situation the system is not at thermal equilibrium when $\beta=0$, and a positive current flows from the hot to the cold reservoir.  
The interesting feature here is that, changing $\beta$ allows to reverse the direction of the currents, so that the system operates as a heat pump \cite{borlenghi14b}. Note also that the main effect of the gradient is a \emph{phase shift} of the current, 
which is the same effect obtained by changing $\beta$.
The difference $p_1-p_2$ between the oscillator powers is displayed in Fig.\ref{fig:figure3} (c). It represents the variation of oscillator amplitudes $\Delta p=p1-p2$ (or "particle numbers" in the BEC and DNLS language \cite{trombettoni01,iubini12}) as a 
function of $\beta$.

Fig.{\ref{fig:figure3}}(d) reports $j^M$ vs $\Delta T=T_1-T_2$, for different values of $\beta$. The computations were performed starting from thermal equilibrium, with $T_1 = T_2 = T_0$. Then, we have increased one of the two temperatures at a time, while keeping the other fixed at $T_0=0.2$. 
The black circles corresponds to the case $\beta=0$, and display the usual rectification effect for heat and spin rectifiers, due to stochastic phase synchronisation \cite{casati04,lepri11,borlenghi14a,borlenghi14b,ren13,ren14} . Here the current increases only for $\Delta T>0$ and remains close to 0 when $\Delta T<0$. 
However, note that the profiles change dramatically with $\beta$. The case $\beta=3\pi/2$ (green diamonds) shows a downwards shift of the current, which now remains constant at negative $\Delta T$. 
The more striking feature occurs when $\beta=\pi/2$ and the current is reversed (red squares), becoming negative at positive $\Delta T$. 
This is consistent with the model, since between the two cases there is a phase difference of $\pi$, which corresponds to a change of sign in the current.

We conclude by discussing phase-controlled spin transport in a realistic setup. The double disk nano-pillar shown in Fig.\ref{fig:figure1}(b) was studied numerically with the Nmag \cite{fischbacher07} micromagnetic software.
The system consists of two Py nano-disks with radius $R=20$ nm and thicknesses $t_1=5$ nm, $t_2=3$ nm separated by a 4 nm spacer. The applied fields $|\bm{H}|=1$ T along $z$ defines the precession axis of the
magnetisation. All the parameters are the same used in Ref. \cite{borlenghi14b}, to which we refer for a thorough discussion.

The nano-pillar is modeled with finite element mesh and the dynamics is computed by solving the Landau-Lifschitz-Gilbert equation \cite{gurevich96} at each mesh node. 
The output consists of the two magnetisation vectors ${M_n(\bm{r}_n,t)}$, which depend on the spacial coordinate $\bm{r}_n$ inside disk $n=1,2$.
The dynamics of the bi-layer system is much more complex than Eqs. (2,3). Although several coarse-graining procedures allow to reduce it to a DNLS-like evolution 
\cite{slavin09,borlenghi14b}, micromagnetic simulations do not offer the possibility to directly control the phase of the coupling between the two disks. 

For this reason the external driving fields $f_1=h{\rm{e}}^{i\omega t}$ and $f_2=h{\rm{e}}^{(i\omega t+\beta)}$ need to be introduced to force the dynamics and control the phase mismatch, see Fig.\ref{fig:figure1}(b). 
Those fields, which are circularly polarised in the $xy$ plane, have intensity $h=0.015$ (with respect to the unit magnetisation vector), frequency $\omega=14.5$ GHz and a phase difference $\beta$.

The collective dynamics, given by the volume average ${\bm{M}}_n(t)=\int_{V_n}{\bm{M}}_n(\bm{r}_n,t){\rm{d}}^3\bm{r}_n/V_n$, 
consists of two classical macrospins $\bm{M}_n(t)$ precessing around $z$. In terms of the stereographic variables $\psi_n=(M_n^x+iM_n^y)/(1+M_n^z)$, the equation of motion for the collective dynamics has
the form of a Schr\"odinger  dimer \cite{slavin09,borlenghi14b}. 

The simulations where performed at zero temperature. The system was evolved for 15 ns, with a time step of 1 ps.
The relevant observables, computed after a transient of 5 ns, were computed as a function of $\beta$.

\begin{figure}
\begin{center}
\includegraphics[width=0.99\columnwidth]{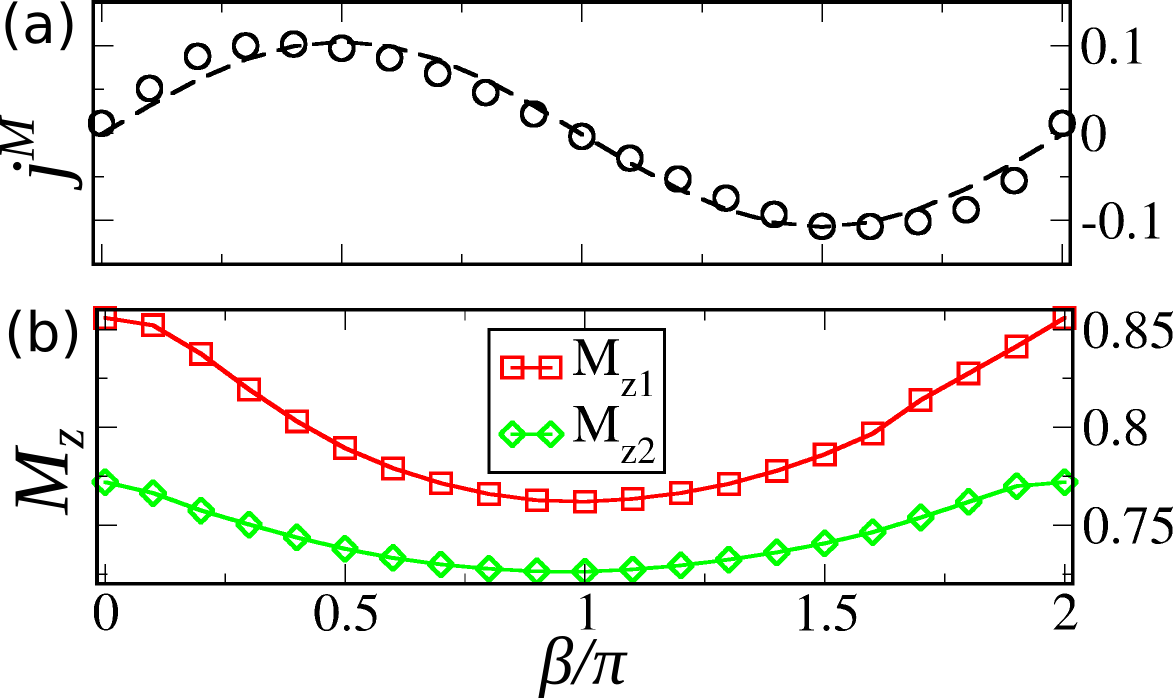}
\end{center}
\caption{(Color online) Micromagnetic simulations. (a) Magnetisation current $j^M$  vs $\beta$, fitted with the function $0.1\sin\beta$ (dashed lines). (b) Dependence of $M^z_{1,2}$ on $\beta$, which accounts for the transfer of magnetic moment between the disks. The line is a guide to the eye.}
\label{fig:figure4}
\end{figure}

The macrospins have the same phase mismatch as the driving fields. The magnetisation current $j^M$, shown in Fig.\ref{fig:figure4}(a), is proportional to $\sin\beta$ as discussed in the model.
The variation of the $z$ components of the magnetisation, displayed in Fig.\ref{fig:figure4}(b) shows that the phase mismatch controls the net transfer of spin between the two disks. 
 In spin valves nano-pillars, this feature could be applied to the fast control of the electrical resistivity.


In conclusion, the transport phenomena described here appear as general properties of classical systems out of thermal equilibrium,
and could be observed in a large class of systems, including mechanical oscillators, optical and nanophononics devices. 

In spin-valve nano-pillars, this effect could be used to control the resistivity through the giant magneto resistance effect \cite{baibich88} or for spin pumping \cite{tserkovnyak02}. Other possible applications include the propagation of spin wave currents in wave guides and the control of domain wall motion \cite{yamaguchi04}.

We acknowledge financial support from the Swedish Research Council (VR), Energimyndigheten (STEM), the Knut and Alice Wallenberg Foundation, the Carl Tryggers Foundation, the Swedish e-Science Research Centre (SeRC) and the Swedish Foundation for Strategic Research (SSF). S.I. acknowledges financial support from the EU-FP7 project PAPETS (GA 323901). We thank the hospitality of the Galileo Galilei Institute for Theoretical Physics, during the 2014 workshop "Advances in Nonequilibrium Statistical mechanics", where part of this work was performed.

\end{document}